\documentclass[aps,a4paper,reprint,showpacs,amsmath,amssymb,superscriptaddress,floatfix,footinbib]{revtex4-2}

\usepackage{amssymb}
\usepackage{latexsym}
\usepackage[dvips]{graphicx}
\usepackage{epsfig}

\usepackage{dcolumn}
\usepackage{amsmath}
\usepackage{amsfonts}
\usepackage{bm}
\usepackage{color}

\usepackage[normalem]{ulem}
\usepackage{hyperref}
\hypersetup{colorlinks,linkcolor=blue,urlcolor=blue,citecolor=blue}

\newcommand{\be}{\begin{equation}}
\newcommand{\ee}{\end{equation}}
\newcommand{\bea}{\begin{eqnarray}}
\newcommand{\eea}{\end{eqnarray}}

\newcommand{\s}{\sigma}

\begin{document}

\title{Suppression of ballistic helical transport by isotropic dynamical magnetic impurities}

\author{Oleg M. Yevtushenko$^*$}
\affiliation{Institut f{\"u}r Theorie der Kondensierten Materie,
             Karlsruhe Institute of Technology, 76128 Karlsruhe, Germany}

\author{Vladimir I. Yudson}
\affiliation{Laboratory for Condensed Matter Physics, National Research University Higher School of Economics, Moscow, 101000, Russia}
\affiliation{Russian Quantum Center, Skolkovo, Moscow Region 143025, Russia}

\date{\today }

\begin{abstract}
Dynamical magnetic impurities (MI) are considered as a possible origin
for suppression of the ballistic helical transport on edges of 2D
topological insulators. The MIs provide a spin-flip backscattering of itinerant
helical electrons. Such a backscattering reduces the ballistic conductance if
the exchange interaction between the MI and the electrons is anisotropic and
the Kondo screening is unimportant. It is well-known that the isotropic
MIs do not suppress the helical transport in systems with axial
spin symmetry of the electrons.
We show that, if this symmetry is broken, the isotropic MI acquires an effective
anisotropy and suppresses the helical conductance. The peculiar underlying mechanism
is a successive
backscattering of the electrons which propagate in the same direction and have different
energies. The respective correction to the linear conductance
is determined by the allowed phase space of the electrons and scales with temperature
as $ T^4 $.  Hence, it disappears at small temperatures.
This qualitatively distinguishes effects governed by the MIs with the induced and
bare anisotropy; the latter is the temperature independent.
If $ T $ is smaller than the applied bias, finite $ e V $, the allowed phase space
is provided by the bias and the differential conductance scales as $ (e V)^4 $.
%%%
% We compare effects of the broken spin symmetry
% combined with either the MI or with electron interactions. We also
% discuss an interplay between the broken spin symmetry and the
% anisotropy of the MIs and resulting unusual temperature dependence
% of the helical conductance.
%%%
\end{abstract}

\maketitle

\section*{Introduction}

Suppression of the edge transport in two-dimensional (2D) topological insulators
\cite{Molenkamp-2007,konig_2013,EdgeTransport-Exp1,EdgeTransport-Exp2}
has attracted considerable attention of the community and remains the hot
topic even after a decade of the
intensive research because of the absence of the fully self-consistent theoretical
explanation. The time-reversal symmetry and the non-trivial topology of the bulk
guarantee helicity (lock-in relation between spin and direction of propagation)
of the gapless one dimensional (1D) edge modes \cite{HasanKane,QiZhang,TI-Shen}.
Helicity prohibits an elastic single-particle backscattering by a spinless potential.
%%%
% cannot
%%%
% and spin-preserving elastic scattering, which converts
%%%
% convert the helical electron into its counterpart from the
% Kramers doublet.
%%%
Thus, at least in the absence of interactions,
the helical modes are not liable to effects of material imperfections,
e.g. localization. Many physical mechanisms, which
are beyond the simplest single-particle picture and can suppress the helical
conductance, have been suggested aiming to explain experimental
data: multi-particle backscattering \cite{WuBernevigZhang,Oreg-MPBS},
electron-electron/phonon interactions and inelastic scattering of the helical electrons
\cite{schmidt_2012,vayrynen_2013,Mirlin-HLL,vayrynen_2014,vayrynen_noise-2018,groenendijk_2018,mcginley_2021},
and their exchange interaction with nanomagnets \cite{meng_2014,novelli_2019}
or localized dynamical magnetic moments -- magnetic impurities (MI),
\cite{MaciejkoLattice,CheiGlaz,AAY,YeWYuA,kimme_2016,vayrynen_2016,vayrynen_2017,Klinovaja_Loss_2017,RKKY-HLL-Bulk,Yevt_2018,hsu_2018,zheng_2018}
to name just a few. We focus on the latter mechanism below.
%%%
% and use there the term (in)elastic scattering of the electrons to denote
% those processes where the electron kinetic energy in (not)conserved.
%%%

An exchange interaction of the helical electrons with the MIs can result in energy-preserving
backscattering accompanied by the spin-flip. That is why the MIs were considered
as a serious obstacle for the ballistic helical transport in the topological insulators
even with the spin axial symmetry of the electrons
%%%
% in setups where the Kondo screening can be neglected , e.g. in the broad range of
% temperatures above the Kondo temperature
%%%
\cite{MaciejkoOregZhang}. As a matter of fact, the MIs, which are isotropically
coupled to the itinerant electrons (the isotropic MIs for brevity), cannot themselves break
the spin U(1) symmetry and, therefore, cannot influence the dc conductance \cite{FurusakiMatveev},
see also Refs. \cite{AAY,YeWYuA,TI-SupSol,OYeVYu_2019}.
The anisotropically coupled MIs (the anisotropic MIs for brevity)
violate the spin conservation and are able to suppress the ballistic
conductance but only if they are not Kondo screened. This requires either a high density
of the MI array \cite{AAY,YeWYuA,TI-SupSol,OYeVYu_2019}, where the Kondo effect is overwhelmed
by the MI correlations \cite{doniach_1977}, or the temperature / the applied bias being larger
than the Kondo temperature, $ {\rm max} \{ T, (e V) \} > T_K $, or a large value of the MI spin,
$ S > 1/2 $ \cite{vinkler_2020,kurilovich_2017,TI-Anisotr-KI}.

The situation is crucially different in the topological insulators without axial spin symmetry of the
electrons \cite{schmidt_2012,Mirlin-HLL,vayrynen_noise-2018}.
The absence of the spin symmetry can be caused, e.g., by the Rashba spin-orbit interaction,
and does not contradict to the concept of the helical modes on the edges of the topological insulators
\cite{KaneMeleQSH}. In these systems, the helical electrons acquire a dependence of the spin orientation
on their energy, the so-called spin texturing (ST) \cite{rod_2015}.
%%%
% The ST leads to the violation of the spin conservation and,
%%%
% It is known that, if the conduction electrons with the different energies participate in the
% backscattering, the helical conductance can be suppressed.
%%%
The well-known effect of the ST is the suppression of the helical conductance caused by the
inelastic backscattering of the interacting electrons \cite{schmidt_2012,Mirlin-HLL}.
Another effect of the  ST, which
%%%
% is rather simple but
%%%
has not been addressed in the literature, is
an influence of the isotropic (unscreened) MI on the helical transport of the noninteracting electrons.
Since the scattering of a given electron by the MI does not change the electron energy,
one could surmise that the ST is unimportant and the isotropic MI could not have any effect
on the linear helical conductance similar to the model with spin axial symmetry \cite{FurusakiMatveev}.
This guess holds true but only at zero temperature.

{\it We show in the present paper} that, even in the absence of the electron interactions,
the isotropic (unscreened) MI yields negative corrections to the dc differential helical conductance
at finite temperatures, $ e V < T $, or at a finite bias, $ T < eV $. The underlying mechanism
%%%
% for suppression of the ballistic conductance by the isotropic MI
%%%
is an energy-preserving successive backscattering of the helical electrons which propagate
in the same direction and have different energies. We elaborate this mechanism at
qualitative and quantitative levels, derive the ST correction to the helical conductance, which scales in
the proper limiting cases as $ \delta G \propto {\rm max} \{ T^4, (eV)^4 \} $,
and compare our results with the suppression of the conductance caused by the anisotropic MI
and by the combined effects of the ST and the electron interactions.

We would like to keep the discussion at the most transparent level and, therefore,
use the representative example of the spin-1/2 MI weakly coupled to the non-interacting
helical electrons at temperatures above $ T_K $. The derivation of the differential
conductance in this setup suffices to explain the combined effect of the MIs and
the ST on helical transport. We touch upon more complicated situations, including
Kosterlitz-Thouless-like renormalizations of the MI-electrons coupling, very briefly
and at a qualitative level.

\begin{figure}[t]
\begin{center}
   \includegraphics[width=0.48 \textwidth]{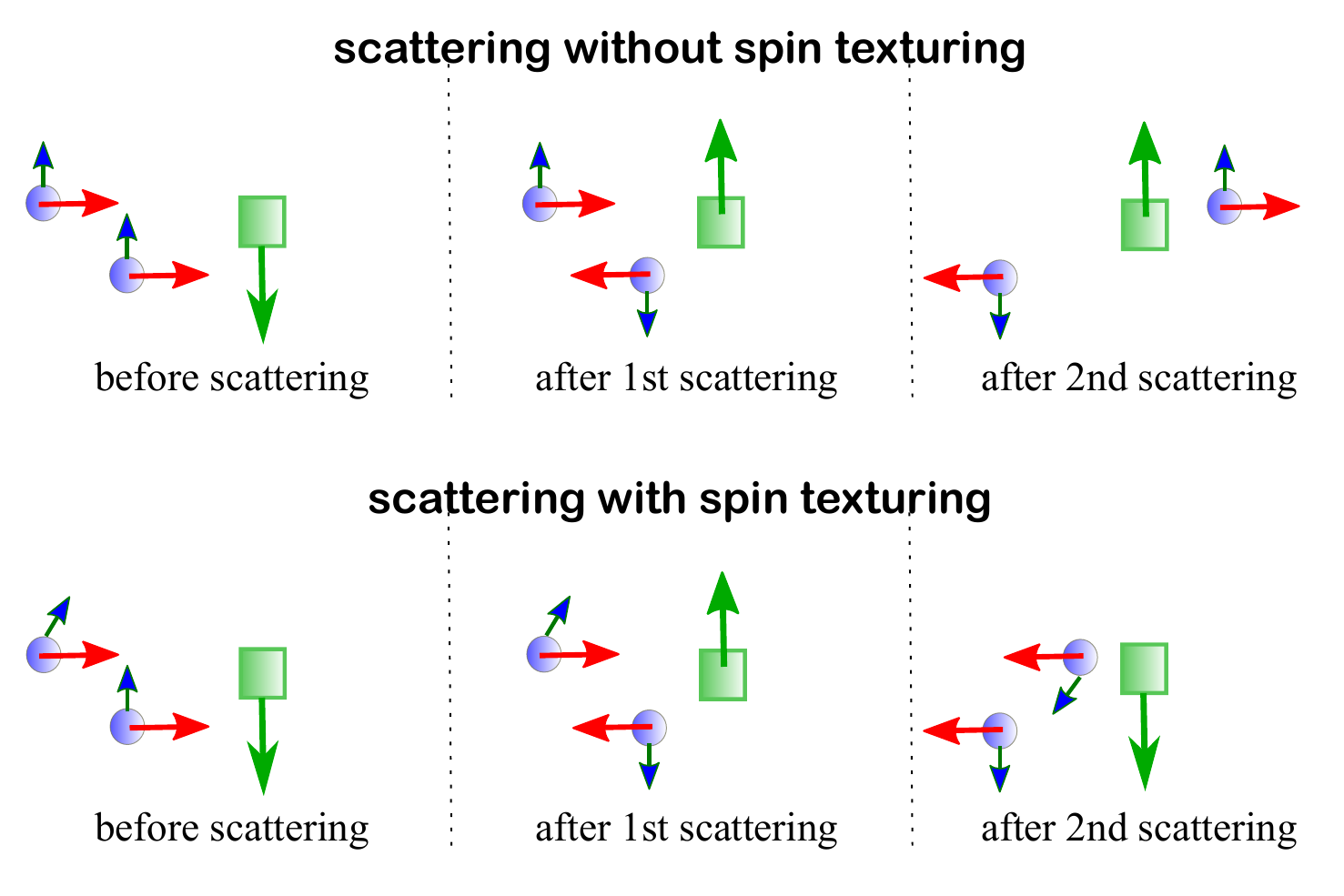}
\end{center}
\vspace{-0.5 cm}
   \caption{
        \label{Backscattering}
        Successive scattering of two right moving helical electrons (shaded circles),
        which have different energies,
        by the MI (shaded square), see the explanation in the text.
           }
\end{figure}

\section*{Qualitative discussion}

The succesive backscattering of the helical electrons by the isotropic MI is illustrated by
Fig.\ref{Backscattering} where the cases with and without the ST are compared.
Let us consider, for example, scattering of two right-moving helical electrons,
see left panels of Fig.\ref{Backscattering}.

If there is no ST, the MI spin becomes parallel to that of all incoming right-moving electrons
already after the first backscattering. This blocks consecutive spin-flips and, therefore, the MI is unable to
backscatter one after another two electrons, see the upper panel of Fig.\ref{Backscattering}.
Two electrons can be backscattered by the MI successively only if they have different chirality and,
as a result, the dc helical conductance is not changed \cite{FurusakiMatveev}. It is a well-known
example of systems where transport is ballistic despite backscattering of the individual electrons.

If, on the contrary, there is the ST, the first backscattering aligns the MI spin with that of the
first incoming electrons before it is backscattered.
%%%
% backscattered electron on a given energy.
%%%
Due to the ST, incoming electrons with another energy have different orientation of the spin.
%%%
% and can be successively backscattered by the MI.
%%%
Therefore, the MI can backscatter two helical electrons of the same chirality one after another
(see the lower panel of Fig.\ref{Backscattering}) which results in the suppression
of the ballistic helical conductance.
%%%
% and is the direct consequence of the absence of the spin conservation.
%%%
This requires a finite energy shell of partially filled electron states around the Fermi energy.
In the linear response regime, the allowed phase space of the electrons decreases with lowering
temperature and the probability of the above described successive backscattering vanishes in the
limit of small $ T $.

Before presenting a more rigorous description, the scaling of the MI governed correction to the
dc conductance can be deduced from a phenomenological approach. Expressions for the backscattering
current and for the respective correction to the differential conductance read as follows:
\begin{equation}\label{Current}
  {\cal J} = e \partial_t \left( N_R - N_L \right) = 2 e \partial_t N_R , \
  \delta G = \partial_V {\cal J} ;
\end{equation}
where $ e $ and $ N_{R,L} $ are the electron charge and numbers of the right/left moving chiral
electrons, respectively and we have taken into account the particle conservation law which relates
chiral currents, $ {\cal J}_R = - {\cal J}_L $. As we have already mentioned in Introduction,
the backscattering caused by the isotropic MI cannot lead to the backscattering current if the
spin axial symmetry is present. Breaking this symmetry leads to the ST and to the effective
anisotropy of the MI. In the case of a weak ST, the anisotropic part of the coupling between the
MI and the itinerant electrons is $ \sim k^2 $, with $ k $ being the electron momentum \cite{schmidt_2012}.
The corresponding (leading in the electron-MI coupling constant)
contribution to $ {\cal J} \propto \partial_t N_R $ is governed by the
product of the external voltage multiplied by the square of the anisotropic part of the coupling
constant, $ {\cal J} \sim e V \times k^4 $. The typical value of $ k $ can be estimated
as $ k \sim {\rm max} \{ T / v_F , e V / v_F \} $. Hence, one may expect $ \delta G \sim
T^4 {\cal F}(e V / T) $, which is reduced to
\begin{equation}\label{Scaling}
 \delta G \propto {\rm max} \{ T^4, (eV)^4 \} ,
\end{equation}
in corresponding limiting cases.
%%%
% The backscattering rate can be found from the
% Fermi golden rule, which involves the momentum integral of the square of the anisotropic part of
% the coupling between the MIs and the itinerant electrons multiplied by corresponding Pauli
% factors. If the anisotropy is governed only by the weak ST, the square of the coupling
% constants is $ \sim k^4 $, with $ k $ being the electron momentum \cite{schmidt_2012}. In
% the linear response regime, the phase space of the momentum integral is limited by the thermal
% volume, $ k \sim T / v_F $. Thus, the power counting yields
% \[
%  \delta G (V \to 0) \propto T^4 .
% \]
%
% In the opposite limit $ T \ll e V $, the phase space is provided by the voltage and one can
% expect a crossover from the $ T $-scaling to the $ e V $-scaling:
% \[
%  \delta G ( T \ll e V ) \propto (eV)^4 .
% \]
%%%
Such an argumentation suggesting the scaling Eq.(\ref{Scaling}) is rather general
and does not depend on the value of the MI spin at $ {\rm max} \{ T, eV \} \gg T_K $. On the other
hand, the above phenomenology is based on the intuition suggested by the Boltzmann kinetic equation, whose
validity is a priory not clear. The kinetic equation must include distribution functions of the
helical electrons and of the MI. The latter can be used only if the MI density matrix is diagonal.
This is generically not correct in arbitrarily chosen basis \cite{kurilovich_2017} while the
a proper basis is not know in advance, see calculations below. Besides,
the power counting is unable to predict the scaling function $ {\cal F}(e V / T) $
%%%
% coefficients in front of the scaling function, which
% can be equal to zero due to some cancellations, while the full description of
%%%
describing the crossover regime, $ T \sim e V $, which might be of interest for comparing with
experimental results. Hence, one needs a more rigorous theory of the backscattering current governed
by the MI. We develop it in the next section by using the master
equation approach for our main example of the spin-1/2 MI.

\section*{Model and method}

Massless 1D fermions propagating on the edge of the 2D topological
insulators are described by the standard Dirac Hamiltonian
\be
  \label{H0}
  H_0 = v_F \sum_k k \left( \Psi^\dagger_{\rm ch}(k) \hat{\sigma}^z \Psi_{\rm ch}(k) \right) .
\ee
Here $ \Psi^{\rm T}_{\rm ch} = \{ \psi_{R}, \psi_{L} \} , \hat{\sigma}^z \equiv {\rm diag} (1, -1) $,
$ \psi_{R/L} $ are chiral fermionic operators of right/left moving modes, $ v_F $ is the Fermi velocity.
The Hamiltonian of the exchange interaction between the
fermions and the MI is naturally written in the spin fermionic basis, $ \Psi^{\rm T}_{\rm sp} = \{
\psi_{\uparrow}, \psi_{\downarrow} \} $:
\be
  \label{Hint}
  H_{\rm int} = \sum_{j} \sum_{k_{1,2}} J_j \Psi^\dagger_{\rm sp}(k_1) \hat{S}^j \hat{\s}^j \Psi_{\rm sp}(k_2), \
  j = x, y, z;
\ee
with $ \hat{S}^j $ and $ \hat{\s}^j $ being the localized spin-1/2 operator and the Pauli
matrices, respectively. The MI is located at position $ x = 0 $. For our purposes, it
suffices to assume the diagonal coupling matrix with small diagonal entries, $ J_j \ll  v_F $.
%%%
% , and to develop the perturbation theory up to the second order in $ J_j $.
%%%
The spin and chiral bases are related by the $ k $-dependent rotation \cite{schmidt_2012,dolcetto_2016}
\begin{equation}\label{Bmatr}
  \Psi_{\rm sp}(k) = \hat{B}(k) \Psi_{\rm ch}(k), \ \hat{B}(k) \equiv \cos(\theta(k)) +
                                           i \hat{\s}^y \sin(\theta(k)) .
\end{equation}
We have introduced the angle of the spin rotation caused by the ST: $ \theta(k) \equiv
(k / k_0)^2 $. The ST is typically weak, $ k / k_0 \sim {\rm max} \{ T, eV \}  / k_0 v_F \ll 1 $.
%%%
% , with $ D $ being the UV energy scale, e.g., the bulk gap of the topological insulator.
%%%
However, using Eq.(\ref{Bmatr}) [without an expansion of $ \hat{B} $ in powers of $ \theta(k) $]
is technically more convenient. Eq.(\ref{Hint}) in the chiral basis reads:
\bea
  \label{Hint-ch}
  H_{\rm int} & = & \sum_{k_{1,2}} \Psi^\dagger_{\rm ch}(k_1) \hat{{\cal G}}(k_1,k_2) \Psi_{\rm ch}(k_2), \\
  \label{G_k}
  \hat{{\cal G}}(k_1,k_2) & = & \sum_j J_j \hat{S}^j \left( \hat{B}^\dagger(k_1) \hat{\s}^j \hat{B}(k_2) \right) \, .
%%%
%  \nonumber
%%%
\eea
Eq.(\ref{G_k}) shows that the ST-caused rotation by the matrix $ \hat{B} $ changes (or
even induces in the isotropic case) the anisotropy of the coupling.
%%%
% constants $ J_{x,y,z} $.
%%%
This effective anisotropy depends on the electron momentum and gives rise to the suppression
of the helical conductance.

The expression for the number of the chiral electrons, which enters Eq.(\ref{Current}) for the
backscattering current, reads:
\begin{equation}\label{PartNum}
%%%
%  & &
%  {\cal J} = e \partial_t \left( N_R - N_L \right) = 2 e \partial_t N_R , \\
%  & &
%%%
  N_{\alpha} = \sum_k {\rm Tr} _{e,MI}\{ \psi^\dagger_\alpha(k) \psi_\alpha(k) \hat{\rho}(t) \} , \
  \alpha = R, L.
\end{equation}
Trace in Eq.(\ref{PartNum}) is calculated with respect to
the helical electrons, $ {\rm Tr}_e $, and the MI, $ {\rm Tr}_{MI} $; $ \hat{\rho}(t) $ denotes
%%%
% The expectation value of the chiral particle number operator, $ N_{R/L} $, is calculated
% with the help
%%%
the total (describing the electrons and the MI) density matrix in the interaction representation
with respect to $ H_0 $.
In the standard approach, which involves
the Markovian approximation (see, e.g., the textbook \cite{le_bellac_quantum_2013}),
$ \hat{\rho}(t) $ obeys the evolution equation $ \partial_t \hat{\rho}
= -i [{\cal H}_{\rm int} (t), \hat{\rho}(-\infty)] + {\cal L}[\hat{\rho}(t)] $, where
\begin{eqnarray}\label{ME_rho}
    {\cal L}[\hat{\rho}(t)]  & \equiv &
        - \int_{-\infty}^t \, {\rm d} t' \,
        \left[ {\cal H}_{\rm int} (t),
            \left[ {\cal H}_{\rm int} (t'), \hat{\rho}(t)
            \right]
        \right] ,
\end{eqnarray}
%%%
% With accuracy of order $ O(J_{x,y,z}^2) $,
%%%
and, in the second order in $ H_{\rm int} $,
$ \hat{\rho} $ in Eq.(\ref{ME_rho}) can be factorized $ \hat{\rho} \simeq \hat{\rho}_e \hat{\rho}_{MI} $.
Here $ \hat{\rho}_e $ is the density matrix of the free electrons coupled to the leads, $ \hat{\rho}_{MI} \equiv
{\rm Tr}_e \{ \hat{\rho} \} $ is the reduced density matrix of the MI.
%%%
% $ {\rm Tr}_e $ denotes the trace over the fermionic degrees of freedom;
%%%
The latter matrix obeys the master equation  $ \partial_t \hat{\rho}_{MI} =
{\rm Tr}_e \left\{ {\cal L}[\hat{\rho}_e
\hat{\rho}_{MI}(t)] \right\}$ (the linear in $ {\cal H}_{\rm int} $ term
does not contribute to $ {\rm Tr}_e $).

The effect which we study is governed by real processes. Virtual processes lead to
renormalizations, which are unimportant at $ {\rm max} \{ T, eV \} \gg T_K $. Therefore,
we neglect the virtual processes and focus on the energy-preserving scattering of the
electrons by the MI.
%%%
% For the problem under consideration,
%%%
The master equation takes the following form
\bea
\label{ME_rho-KI}
   & &
    \partial_t \hat{\rho}_{MI} = - \frac{\pi}{(2 \pi v_F)^2} \sum_{\alpha,\beta}
                      \int \!\! {\rm d} \, \epsilon \Bigl\{
                              f_e^\alpha(\epsilon) \left[ 1 - f_e^\beta(\epsilon) \right]  \\
%%%
%    {\rm Tr}_e \left\{ {\cal L}[\hat{\rho}_e \hat{\rho}_{MI}(t)] \right\}.
%%%
    & &
    \times \left[
        \{ \hat{\cal G}_{\alpha\beta}(\epsilon) \hat{\cal G}_{\beta\alpha}(\epsilon), \hat{\rho}_{MI} \}_+ -
      2 \hat{\cal G}_{\beta\alpha}(\epsilon) \hat{\rho}_{MI} \hat{\cal G}_{\alpha\beta}(\epsilon)
           \right]
       \Bigr\} .
    \nonumber
\eea
Here $ \alpha, \beta = R, L $; $ \hat{\cal G}(\epsilon) \equiv \hat{\cal G}(\epsilon/v_F, \epsilon/v_F ) $;
$ f_e^{R/L} $ are fermionic distribution functions:
\be
  f_e^{R/L}(\epsilon) = \frac{1}{\exp\left[ ( \epsilon - \mu_{R/L} ) / T \right] + 1};
\ee
$ \mu_{R/L} $ are chemical potentials of the corresponding leads; we assume $ \mu_{R/L} = \pm e V /2
\Rightarrow \delta \mu \equiv \mu_R - \mu_L = e V $.
The stationary solution of Eq.(\ref{ME_rho-KI}), $ \hat{\rho}_{MI}^{\rm st} $, determines the backscattering
dc current:
% \be
%   \label{StatCurr}
$
   {\cal J}_{\rm dc} = 2 e {\rm Tr}_{e,MI} \left\{
     \psi^\dagger_R \psi_R \,
     {\cal L}[ \hat{\rho}_e \hat{\rho}_{MI}^{\rm st} ]
                                                  \right\}
$.
% \ee
Similar to Eq.(\ref{ME_rho-KI}), we find
\bea
   \label{ME_Jdc}
   {\cal J}_{\rm dc} & = & \frac{4 \pi e}{(2 \pi v_F)^2} \int {\rm d} \epsilon \, \\
                     &   &
     \Bigl(
     f_e^L(\epsilon) \left[ 1 - f_e^R (\epsilon) \right]
     {\rm Tr}_{MI} \{ \hat{\cal G}_{LR}(\epsilon) \hat{\cal G}_{RL}(\epsilon) \hat{\rho}_{MI}^{\rm st}\} - \cr
                     &   &
   - f_e^R(\epsilon) \left[ 1 - f_e^L (\epsilon) \right]
     {\rm Tr}_{MI} \{ \hat{\cal G}_{RL}(\epsilon) \hat{\cal G}_{LR}(\epsilon) \hat{\rho}_{MI}^{\rm st}\}
     \Bigl) .
\nonumber
\eea
%%%
% The correction to the linear dc conductance is defined by the standard relation $ \delta G = \lim_{\delta \mu \to 0}
% \left( e {\cal J}_{\rm dc} / \delta \mu \right), \ \delta \mu \equiv \mu_R - \mu_L = e V $, where $ V $ is the applied
% voltage.
%%%

\section*{Results and their discussion}

\subsection*{Linear response}

Let us first study the linear response, $ V \to 0 $, and find the temperature-dependent correction to
dc linear conductance, $ \delta G_L(T) \equiv \delta G (T, V=0) $.

After a straightforward algebra, we obtain the following answers for the stationary density matrix
of the MI:
\begin{eqnarray}
  \label{RhoStAnswer}
  \hat{\rho}_{MI}^{\rm st} & = & \frac{1}{2} + \frac{\delta\mu}{T} \tilde{J}_y\times \\
                           & & \times \left( \frac{ \tilde{J}_x I_c({\cal T}) }{\tilde{J}_x^2 + \tilde{J}_y^2} \hat{S}^z
                                           - \frac{ \tilde{J}_z I_s({\cal T}) }{\tilde{J}_z^2 + \tilde{J}_y^2} \hat{S}^x \right)
                                      + o(\delta\mu) ;
  \nonumber
%                                      \cr
\end{eqnarray}
and for $ \delta G_L $:
%%%
% \begin{eqnarray}
%   \label{CondAnsw}
%   \frac{\delta G}{G_0} & = &
%    - \frac{1}{2} \left( \frac{\tilde{J}_x^2 + \tilde{J}_z^2}{2} + \tilde{J}_y ^2\right)
%    - \frac{ \tilde{J}_x^2 - \tilde{J}_z^2 }{4} I_c(2 {\cal T}) +
%             \\
%            & &
%    + 2 \tilde{J}_y^2
%      \left[ \frac{ \left( \tilde{J}_x I_c({\cal T}) \right)^2 }{\tilde{J}_x^2 + \tilde{J}_y^2} +
%             \frac{ \left( \tilde{J}_z I_s({\cal T}) \right)^2 }{\tilde{J}_z^2 + \tilde{J}_y^2} \right] .
%    \nonumber
% %   \cr
% \end{eqnarray}
%%%
\begin{eqnarray}
  \label{CondAnsw}
  \frac{\delta G_L}{G_0} & = &
   - \frac{1}{2} \frac{ \left( J_x^2 - J_y^2 \right)^2}{J_x^2 + J_y^2}
   + \frac{ \tilde{J}_x^2 - \tilde{J}_z^2 }{4} [ 1 - I_c(2 {\cal T}) ] +
            \\
           & &
   + 2 \tilde{J}_y^2
     \left[ \frac{ \tilde{J}^2_x }{\tilde{J}_x^2 + \tilde{J}_y^2} \left( I^2_c({\cal T}) - 1 \right) +
            \frac{ \tilde{J}^2_z }{\tilde{J}_z^2 + \tilde{J}_y^2} I^2_s({\cal T})
     \right] .
   \nonumber
%   \cr
\end{eqnarray}

Here $ {\cal T} \equiv 2 \left( 2 T / v_F k_0 \right)^2 ; \ \tilde{J}_{x,y,z} \equiv J_{x,y,z} / v_F $;
$ G_0 = e^2 / h $ is the helical ballistic conductance; and we have introduced two functions.
\[
     I_c(x)  =  \int_0^{\infty} {\rm d} \epsilon
                       \frac{\cos \left( x \epsilon^2 \right)}{\cosh^2 (\epsilon)} , \
     I_s(x)  =  \int_0^{\infty} {\rm d} \epsilon
                       \frac{\sin \left( x \epsilon^2 \right)}{\cosh^2 (\epsilon)} .
%   \nonumber
\]
Since $ \delta G_L $ depends on $ T / v_F k_0 $, its temperature independent part corresponds
to the limit of small temperature and, simultaneously, of the vanishing ST.

If the MI is isotropic, $ \tilde{J}_{x,y,z} = \tilde{J} $, Eqs.(\ref{RhoStAnswer},\ref{CondAnsw})
reduce to
\begin{eqnarray}
  \label{RhoStAnswerIso}
  \hat{\rho}_{MI}^{\rm st} & \simeq & \frac{1}{2} + \frac{\delta\mu}{2T}
                                 \left( I_c({\cal T}) \hat{S}^z - I_s({\cal T}) \hat{S}^x \right) \simeq \\
                           \frac{1}{2} & + & \frac{\delta\mu}{2 T}
                                 \left( \left[ 1- \frac{14}{15} \left( \frac{\pi T}{v_F k_0}\right)^4 \right] \hat{S}^z -
                                        \frac{2}{3} \left( \frac{\pi T}{v_F k_0}\right)^2 \hat{S}^x
                                 \right) \! ;
  \nonumber
%                                      \cr
\end{eqnarray}
\begin{eqnarray}
  \label{CondIso}
  \delta G_L^{\rm (iso)}
           & = & - \tilde{J}^2 \left\{ 1 - \left[ I_c^2({\cal T}) + I_s^2({\cal T}) \right] \right\} G_0 \simeq \\
           & \simeq & - \frac{4\tilde{J}^2 G_0}{45}  \left(\frac{2 \pi  T}{ v_F k_0 } \right)^4 .
           \nonumber
\end{eqnarray}
Eq.(\ref{CondIso})
%%%
% is our main result and
%%%
confirms that the isotropic (unscreened) MI is able to
suppress the helical ballistic conductance in the topological insulators with broken axial spin
symmetry of the electrons.

The exponent of the power-law $ T $-dependence of $ \delta G_L^{\rm (iso)} $ is anticipated from the
phase space-arguments of the above qualitative discussion and, simultaneously, the value
$ \delta G_L^{\rm (iso)} (T / v_F k_0 = 0) = 0 $ agrees
with the prediction of Ref.\cite{FurusakiMatveev} on the vanishing effect of the isotropic
MI in the absence of the ST.
The temperature dependence of $ \delta G_L^{\rm (iso)} \sim
( T / v_F k_0 )^4 $ dominates over that of
a correction to the linear conductance, $ \delta G_{\rm int} $, governed by the combined effect of the
ST and the electron interactions. $ \delta G_{\rm int} $ is parametrically smaller in a clean
system and reads as either $ \delta G_{\rm int} \propto ( T / v_F k_0 )^5 $ at $ k_F = 0 $ or
$ \delta G_{\rm int} \propto ( v_F k_0 / T ) \exp(- v_F k_0 / T) $ at $ v_F k_F \gg T $ \cite{schmidt_2012}.
%%%
% In particular, the correction $ \delta G_{\rm iso} $ vanishes at $ T \to 0 $ as $ ( T / v_F k_0 )^4 $.
%%%%

\begin{figure}[b]
\begin{center}
   \includegraphics[width=0.48 \textwidth]{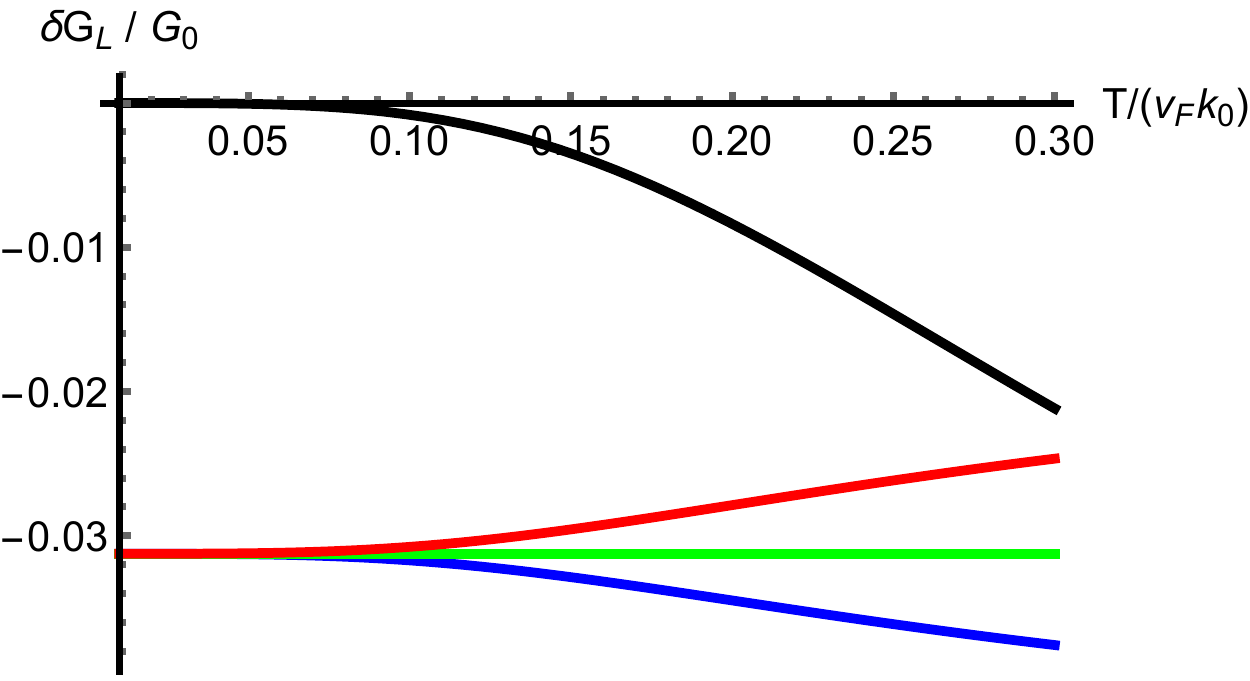}
\end{center}
\vspace{-0.5 cm}
   \caption{
        \label{ConductPlot}
        Temperature dependence of the MI-generated correction to the linear ballistic helical
        conductance for four cases:
        (i)~isotropic MI, $ \tilde{J}_{x,y,z} = 0.25 $ (black curve);
        (ii) anisotropic MI $ \tilde{J}_{x} = 0.25, \tilde{J}_{y} = 0, \tilde{J}_z = 0.35 $ (blue curve);
        (iii) easy xz-plane anisotropic MI, $ \tilde{J}_{x,z} = 0.25, \tilde{J}_{y} = 0 $ (green curve);
        (iv) easy x-axis anisotropic MI, $ \tilde{J}_{x} = 0.25, \tilde{J}_{y,z} = 0 $ (red curve).
           }
\end{figure}

The bare XY-anisotropy of the coupling constants makes $ \delta G_L |_{T \to 0} $ in Eq.(\ref{CondAnsw})
finite (still assuming $ T > T_K $) in agreement with the previous works \cite{vinkler_2020,kurilovich_2017,TI-Anisotr-KI},
see comparison of the isotropic and anisotropic cases in Fig.\ref{ConductPlot}. Curiously,
the ST can decrease the effect of the magnetic anisotropy and make $ G_L ( T = 0) <
G_L ( T \ne 0) $, with $ G_L ( T ) $ certainly being smaller than $ G_0 $. This happens, for instance,
at $ J_y = 0 $ in the range $ 0 < J_z < J_x $, the red curve in Fig.\ref{ConductPlot}. If $ J_y = 0,
J_z = J_x \ne 0 $, the ST does not change $ G_L $ which is finite and temperature independent,
the green curve in Fig.\ref{ConductPlot}. In
the regime $ J_y = 0, J_z > J_x \ne 0 $, the ST enhances the effect of the magnetic anisotropy such that
$ G_L ( T = 0) > G_L ( T \ne 0) $, the blue curve in Fig.\ref{ConductPlot}. Thus, an unusual growth of
the subballistic helical conductance with increasing the temperature might indicate the combined effect
of the anisotropic MI and the ST. The $ T^4 $-dependence of the correction to the conductance can arise
also in a model which describes the combined effect of the ST and the electron interactions in the presence
of a potential disorder; their combined action results in the scaling
$ \delta G_{\rm int+imp} \propto ( T / v_F k_0 )^4 $ at $ v_F / L \ll T \ll v_F k_F $ ($ L $ is the systems size)
\cite{schmidt_2012}. However, this mechanism always leads to the inequality $ G_L ( T = 0) > G_L ( T \ne 0) $.

If one takes into account virtual processes, the coupling constants in Eqs.(\ref{CondAnsw},\ref{CondIso})
acquire an additional temperature dependence due to Kosterlitz-Thouless-like renormalizations
which are very weak and can be safely ignored at $ T \gg T_K $ \cite{MaciejkoOregZhang,MaciejkoLattice}.
Moreover, in a particular case $ J_{x,y} = 0 $, the renormalizations are absent
at the level the first loop of the renormalization group. Therefore, the answer (\ref{CondAnsw}) is
expected to approximately hold true down to ultra-low temperatures in the case of the easy z-axis anisotropy
of the exchange coupling.

\subsection*{Nonlinear response caused by isotropic MI}

\begin{figure}[b]
\begin{center}
   \includegraphics[width=0.48 \textwidth]{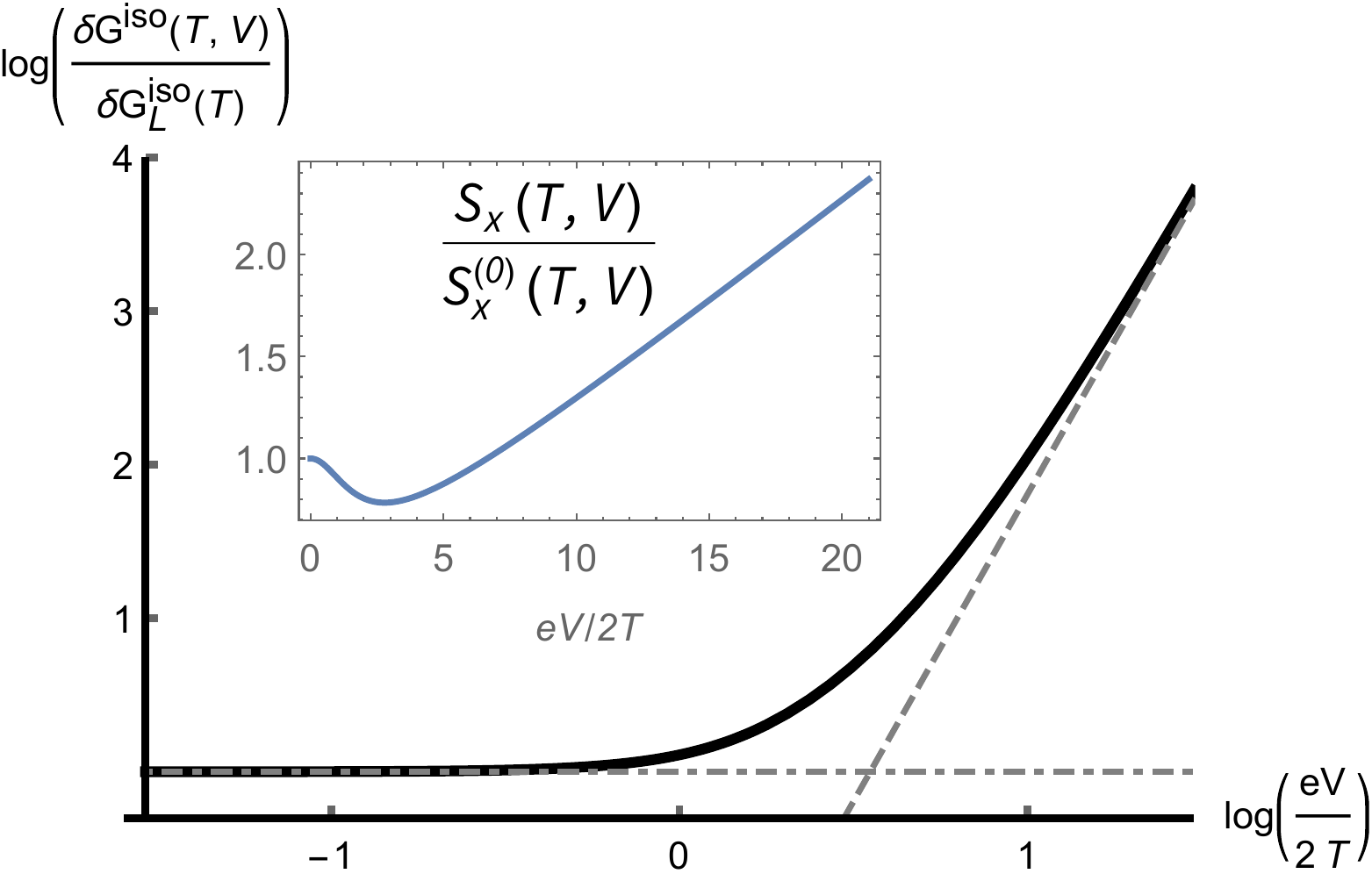}
\end{center}
\vspace{-0.5 cm}
   \caption{
        \label{Nonlin_ConductPlot}
        Main panel: Crossover of the correction to the differential conductance, Eq.(\ref{NonLin_Cross}),
                    between limits of the linear response, $ e V \ll T_K \ll T $ (dash-dotten line), and
                    the finite bias, $ T, T_K \ll e V $ (dashed line).
        Inset: Polarization of the MI spin along $ x $-axis (i.e., along the helical edge), Eq.(\ref{Polarization}).
           }
\end{figure}

Eqs.(\ref{ME_rho-KI},\ref{ME_Jdc}) describe the combined effect of the MI
and the ST on helical transport also at a finite bias. This effect does not vanish
if the temperature approaches zero but the bias is finite,
$ T_K \ll e V $. In this case, the phase
space, which is needed for the successive backscattering of the helical electrons
with the same chirality, is provided by $ e V $ instead of $ T $. It is reflected
by a crossover of the differential conductance from the $ T^4 $-scaling
to the $ (e V)^4 $-scaling. In particular, the correction to the nonlinear
conductance governed by the isotropic MI reads as:

\begin{widetext}
\begin{equation}\label{NonLin_T0}
    \delta G^{\rm (iso)}( T \ll e V ) \simeq
                             - \frac{\tilde{J}^2 G_0}{18} \left( \frac{e V}{v_F k_0} \right)^4 =
                        \frac{5}{8} \left( \frac{u}{\pi} \right)^4 \delta G_L^{\rm (iso)} ; \quad
                        u \equiv \frac{e V}{2T} \!\! .
\end{equation}

The crossover between regimes of the infinitesimal bias, Eq.(\ref{CondIso}),
and the vanishing temperature, Eq.(\ref{NonLin_T0}), is described by a rather
complicated function:
%%%
% \begin{equation}\label{NonLin_Cross}
%    \frac{\delta G_{\rm iso} ( T, V )}{\delta G_{\rm iso} ( T, 0 )} \simeq \frac{2}{(2 \pi)^4}
%     \left\{
%            \pi ^2 \left( 4 \pi ^2 + 15 u^2 \right) \left( 2 - \tanh^2(u) \right) +
%            30 u \left(u^3+\pi ^2 \tanh(u)\right)
%            - 5 u^5 \frac{ 5u + 6\tanh(u) - \frac{u}{\cosh^2(u)} }{[u+\tanh (u)]^2}
%     \right\} .
% \end{equation}
%%%
\begin{equation}\label{NonLin_Cross}
    \frac{\delta G^{\rm (iso)} ( T, V )}{\delta G_L^{\rm (iso)} ( T )} \simeq
       1 - \frac{\tanh^2(u)}{2}
       + \frac{15}{4} \left( \frac{u}{\pi} \right)^2 \left[ 1 + \frac{\tanh(u)}{u} - \frac{\tanh^2(u)}{2} \right]
%%%
%       - \frac{1}{2}\left(1 + \frac{15}{4}\frac{u^2}{\pi^2}\right)\tanh^2{u}
%%%
      +  \frac{5}{4} \left( \frac{u}{\pi} \right)^4
             \left[1 + \frac{4  + 2u \coth(u) - u^2}{2\left(1+ u\coth(u)\right)^2}\right] ;
\end{equation}
see the main panel of Fig.\ref{Nonlin_ConductPlot}.
%%%
% and $ \delta G_{\rm iso} ( T, 0 ) $ is the linear conductance, Eq.(\ref{CondIso}).
%%%
\end{widetext}

Complexity of Eq.(\ref{NonLin_Cross}) at
$ T \simeq e V $ is related to the MI polarization, which develops with increasing
the ratio $ e V / T $. For instance, the partial mean polarization of the isotropic
MI along the helical edge, $ S_x = {\rm Tr} \left[ \hat{S}^x \rho^{\rm st}_{MI}
\right] $, is given by
\begin{equation}\label{Polarization}
    \frac{S_x (T, eV)}{S^{(0)}_x (T, eV)} \simeq
       \frac{ 1 + \left( \frac{u}{\pi} \right)^2 + \left[1 + 3 \left( \frac{u}{\pi} \right)^2 \right] \frac{\tanh (u)}{u}}
       { 1 + u \coth (u) } ;
\end{equation}
see the inset in Fig.\ref{Nonlin_ConductPlot}. Here $ S^{(0)}_x = - (2 u / 3)
( \pi T / v_F k_0 )^2 $ is the corresponding value of the MI polarization in the
linear regime, cf. Eq.(\ref{RhoStAnswerIso}).

\section*{Summary}

We have demonstrated that the spin texturing, which reflects the absence of the
spin axial symmetry of electrons, noticeably changes the influence of dynamical magnetic
impurities on the helical transport on edges of 2D topological insulators.
%%%
% Since the total spin is not conserved
%%%
This is particular pronounced in the emergent ability of the (unscreened) spin U(1)-invariant
MI to suppress the ballistic dc helical conductance. We have exemplified such a suppression
by considering the example of the isotropic MI. This effect does not require the energy transfer
but, nevertheless,
%%%
% Therefore,
%%%
% The backscattering of the non-interacting helical fermions by the MI is elastic and
%%%
% it looks tempting to surmise that the energy dependence of the electron spin orientation
% is irrelevant, and to reduce the physics to the naive picture of the absent ST. If the
% ST were really irrelevant, one would arrive at the conclusion on the insensitivity
% of the ballistic helical conductance to the presence of the isotropic MI. In reality,
% this guess is correct only at zero temperature.
%%%
is temperature-dependent. The corresponding contribution to the linear helical conductance,
$ \delta G_L^{\rm (iso)} $ in Eq.(\ref{CondIso}),
%%%
% , which is governed by the combined effect of the isotropic MI and the ST,
%%%
is negative and its magnitude decreases as $ T^4$
while lowering the temperature. It is related to the underlying physical mechanism of suppression
of the dc conductance: the isotropic MI can backscatter one after another electrons propagating
in the same direction if they have different orientation of spins due to different energies.
If $ T \to 0 $, transport is carried by the electrons on the Fermi
level, the available phase space of the electrons shrinks and the predicted effect
disappears, namely, the isotropic MI does not affect the helical conductance similar
to the systems with the spin axial symmetry (no ST). The predicted $ T^4 $-dependence of $
 \delta G_L^{\rm (iso)} $ is parametrically greater than that anticipated in clean systems with
the ST and the electron interactions \cite{schmidt_2012}.
%%%
% The $ T^4 $-dependence similar to our answers, Eqs.(\ref{CondAnsw},\ref{CondIso}) can appear
% in a more complicated model where a weak potential disorder is added to the ST and the electron
% interactions \cite{schmidt_2012}.
%%%%

If the applied external voltage if finite and $ T \ll e V $, the available phase space is provided
by the voltage and the differential conductance scales as $ G^{\rm (iso)} \propto (e V)^4 $, Eq.(\ref{NonLin_T0}).
The crossover between two scaling regimes is described by Eq.(\ref{NonLin_Cross}). It reflects
a partial polarization of the MI caused by a rather complicated competition of the applied voltage
and the finite temperature.

We note also a curious feature of the combined effect of the ST and the anisotropic MI: if the
bare anisotropy of the coupling between the MI and the conduction electrons is of the easy axis
type (with the axis being directed along the edge of the topological insulator), the ST can effectively
weaken the anisotropy and partially restore the ballisticity of the conductance. This effect could manifest
itself in an unusual growth of the subballistic linear helical conductance with increasing $ T $.

\begin{acknowledgments}
O.M.Ye. acknowledges support from the DFG through the grant YE 157/2-3.
V.I.Yu. acknowledges support from the Basic research program of HSE.
\end{acknowledgments}

\vspace{0.5 cm}

{\it Author contributions}: The authors have made an equal contribution to this paper.

\vspace{0.5 cm}

{\it Competing interests}: The authors declare no competing interests.

\vspace{0.5 cm}

{\it Data availability}: No datasets were generated or analysed during the current study.

\bibliography{Bibliography,TI,Combined-SOI}

\end{document}